# Using Partial Structure R1 to Do Molecular Replacement Calculations

Authors

**Xiaodong Zhang[a]***

[a]Chemistry Department, Tulane University, 6400 Freret Street, New Orleans, Louisiana, 70118, United States

Correspondence email: xzhang2@tulane.edu

**Synopsis**   Applications of the partial structure R1 (pR1) in molecular replacement calculations are demonstrated.

**Abstract**    The concept of partial structure R1 (pR1) is a generalization of the concept of single atom R1 (sR1) (Zhang & Donahue, 2024). The hypothesis is that the deepest hole of a pR1 map determines the orientation and location of a missing fragment. In current implementation, the calculation is divided into two steps. The first step is to detect possible orientations of all missing fragments by the holes of a pR1 map of a free-standing fragment in a 3-dimensional orientation space. The second step is to determine the orientation and location of a missing fragment. To this end, if done strictly, all the candidate orientations are tried. With each candidate orientation, the best choice of location of the missing fragment is determined by the deepest hole of a pR1 map in a 3-dimensional location space. This best choice is combined with the trial orientation to form one candidate orientation-location. After trying all candidate orientations, a list of candidate orientation-locations are formed, from which, the one with the lowest R1 determines the orientation and location of a missing fragment. Then a newer pR1 is defined by including the atoms of this newly determined fragment into the known atoms. This newer pR1 is used to determine the next missing fragment in the same way. To shorten the calculation time, the possible locations of all missing fragments can be predicted by the holes of a pR1 map of a completely disoriented model of a fragment. All these ideas of using pR1 to do molecular replacement calculations have been demonstrated by four example data sets.

## 1. Introduction

Molecular-replacement (MR) (Rossmann & Blow, 1962; Rossmann, 1972, 1990) is a method for phasing an unknown structure by optimally placing a known search molecule or molecules within the target unit cell. Due to wide availability of search molecular models in the Protein Data Bank (PDB), nowadays about 80% of protein structures are solved by MR methods (McCoy *et al.*, 2017). The key step of MR is to determine the orientation and location of a known molecular model within the target unit cell. Originally Patterson targets were used (rotation function, Rossmann & Blow, 1962; translation function, Rossmann *et al.*, 1964). One breakthrough was the recasting of the rotation function in a manner suitable for rapid computation (Crowther, 1972). The traditional R1 was sometimes used as target with some success in translational search (Eventoff *et al.*, 1975; Bott & Sarma, 1976). One drawback of the R1 target is that it is sensitive to scattering factor scaling. The correlation coefficient targets were introduced to overcome this drawback (Harada *et al.*, 1981; Fujinaga & Read, 1987). A more severe drawback of R1 target is that it lacks strict statistical basis. The maximum-likelihood targets address this issue by establishing the calculations on the strict statistical hypothesis testing basis (Bricogne, 1992; Read, 2001; McCoy, 2004). Currently the most sensitive target is the log-likelihood-gain on intensities (LLGI), which is the sum of log-likelihood for individual reflections minus the log-likelihoods of an uninformative model (McCoy *et al.*, 2007; Read & McCoy, 2016). Another probabilistic target based on the joint probability distribution has also been used in an MR program (*REMO*09, Caliandro *et al.,* 2009; Burla, *et al.*, 2020), which can combine with various prior conditions. Currently, it is a general belief that the traditional R1 is not a good target for MR calculations (McCoy, 2017). However, we recently proposed a new concept called the single atom R1 (sR1) and demonstrated that global minimization of sR1's can sequentially locate the missing single atoms (Zhang & Donahue, 2024). This sR1 method is essentially an MR method, in which the known molecules are the single atoms. During our study of the sR1 method, we have also generalized the sR1 concept to a concept called the partial structure R1 (pR1), and a hypothesis has been suggested: the deepest hole of a pR1 map in a 6-dimensional orientation-location space determines the orientation and location of a missing fragment, that is, pR1 can be applied in general MR calculations. The purpose of this report is to demonstrate that pR1 indeed can be applied to general MR calculations.

## 2. Definition of the partial structure R1 (pR1)

The same equations (Zhang & Donahue, 2024) that are used to define the single atom R1 (sR1) are also used to define the partial structure R1 (pR1), among which the core approximation is following equation:

$F_c^2(hkl) \approx [f_1(hkl) \cos 2\pi(hx_1+ky_1+lz_1) + \cdots + f_j(hkl) \cos 2\pi(hx_j+ky_j+lz_j)]^2$

$\qquad + [f_1(hkl) \sin 2\pi(hx_1+ky_1+lz_1) + \cdots + f_j(hkl) \sin 2\pi(hx_j+ky_j+lz_j)]^2$

$$+ f_{j+1}^2(hkl) + \cdots + f_N^2(hkl) \tag{1}$$

When defining sR1, among atoms 1 to j, only j is the single missing atom, while 1 to j-1 are the known atoms. When defining pR1, there are multiple missing atoms: while 1 to i-1 are the known atoms, i to j are a group of missing atoms. Though j+1 to N are also among the missing atoms, their locations do not appear in this equation; on the other hand, the locations of the group of missing atoms i to j do appear in this equation. The group of missing atoms i to j form a fragment, or a partial model, or sometime called a partial structure. The structure of this fragment is known; only its orientation and location are unknown. So, in the end, pR1 depends on the orientation and location of a single missing fragment. If this fragment contains a single atom, pR1 turns into sR1. Though pR1 is a more general concept, usually pR1 refers to a fragment of multiple atoms, while sR1 specifically refers to the single atom situation.

### 3. General considerations of using pR1 to locate missing fragments

Four fragments are studied in this report. The structures of these fragments are shown in Figure 1.

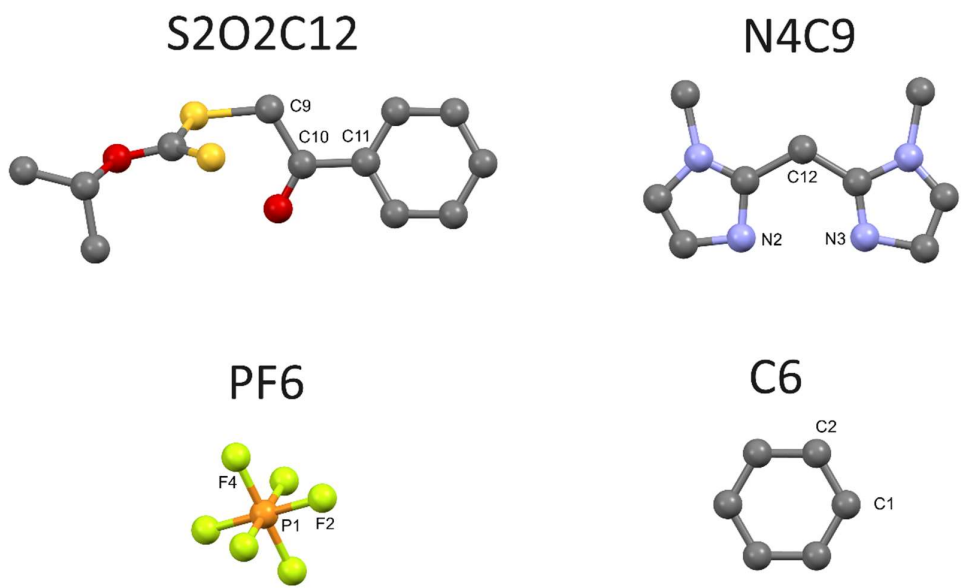

**Figure 1** The structures of four fragments that are studied in this report

A local Cartesian coordinate system is set up for each fragment. The origin of the system sits at the average (weighted by the number of electrons) location of all the atoms in a fragment. The orientation of the system is defined by three unit vectors **x**, **y**, and **z**. To define these unit vectors, two reference vectors **r** and **s** are selected. A reference vector is defined by two reference points: a reference vector is a vector from reference point A to reference point B. The location of an atom, or any other convenient point can be used as a reference point. The unit vectors **x**, **y**, and **z** are related to reference vectors **r** and **s** as follows: **x** = **r**/|**r**|, **y** = (**s**-**s**·**xx**)/| **s**-**s**·**xx** |, **z** = **x**×**y**. Details for selection of reference vectors **r** and **s** for each fragment are deferred to the supporting information.

A Cartesian coordinate system is also set up for the unit cell. The origin of this system overlaps the origin of the unit cell. Its three unit vectors **x**, **y**, and **z** are related to the unit cell vectors **a** and **b** as follows: **x** = **a**/|**a**|, **y** = (**b**-**b**·**xx**)/| **b**-**b**·**xx** |, **z** = **x**×**y**. The Cartesian coordinates of this system and the fractional coordinates of the unit cell are inter-converted during the calculations. When doing translational calculations a grid of 0.4 Å step size is set up within the unit cell.

At start, a fragment is located such that its local Cartesian system overlaps the Cartesian system of the unit cell. The fragment is attached to its local Cartesian system. So, rotation and translation of the fragment is realized by rotating and translating its local Cartesian system in the cell Cartesian system. Translation is realized by translating the local origin in the cell Cartesian system (or rather in the cell fractional coordinate system, which is converted to the cell Cartesian system; mathematical formula describing the relation between fractional coordinates and the cell Cartesian coordinates are deferred to the supporting information). A general rotation is consisted by three simple rotations: a rotation around z-axis (of the local frame) in the direction from x-axis to y-axis through angle ψ; a rotation around x-axis (of the local frame) in the direction from y-axis to z-axis through angle φ; and another rotation around z-axis (of the local frame) in the direction from x-axis to y-axis through angle ζ. Mathematical formula describing a general rotation is deferred to the supporting information.

The space spanned by rotation and translation is a 6-dimensional orientation-location space. (It is 5-dimensional if the fragment is linear.) To coarsely locate the global minimum point or the holes of a pR1 map in this 6-dimensional space, a coarse grid is set up with 0.4 Å step size in translations within the cell and 5-degree step size in rotation angles. The range of rotation angles are: 0 to 360 degrees for ψ, 0 to 180 degrees for φ, and 0 to 360 degrees for ζ. If the fragment has n-fold rotation symmetry, and the rotation axis is arranged along the local z-axis, the range of ζ can reduce to 0 to 360/n. The precision of locating the global minima or the local minimas is refined by halving the step size locally five times.

The general hypothesis is that the deepest hole of a pR1 map in a 6-dimensional orientation-location space determines the orientation and location of a missing fragment (Zhang & Donahue, 2024). Locating pR1 holes in a 6-dimensional space is time-consuming. To cut calculation time, the problem is divided into two 3-dimensional calculations, that is, finding orientation and location in two separate steps.

For a free-standing fragment, that is, a model consisting of a single fragment with no other known atoms, the pR1 only depends on the orientation of the fragment. Therefore, the possible orientations of all missing fragments are detected by the holes in the pR1 map of a free-standing fragment in a 3-dimensional orientation space. Due to possible symmetry of a fragment and the redundancy in representing an orientation with the three rotation angles, many orientation representations are equivalent to each other. Considering two orientation representations, if the atoms of the fragment of one orientation representation can match the atoms of the fragment of the other orientation representation in a one-to-one basis within 0.5 Å, the two representations are considered equivalent. (In this comparison, the types of atoms are ignored. This practice is fine with most fragments. However, in rare cases the type of atoms should be considered, and the algorithm need to be adjusted, with a penalty of longer calculation time.) The non-equivalent orientation representations are filtered out from all detected representations. These serve as the candidate orientations.

The candidate orientations are ranked from smallest R1 to highest R1 and are labelled by 0, 1, 2, etc. In some situations, for example, if it is known that there are only two fragments in the structure, then it is obvious that one fragment has orientation 0 and the other has orientation 1. In such a situation, fragment 0 takes orientation 0, and its location is determined by the deepest hole in a pR1 map of a 3-dimensional location space. After determination of fragment 0, a new pR1 is defined by including the atoms of fragment 0 as known atoms and taking orientation 1 the location of fragment 1 is determined by the deepest hole of this new pR1 map in a 3-dimensional location space. In other situations, there are multiple fragments in the structure. In such situations, to determine a missing fragment, it is necessary to try all possible orientations. For each trial orientation, the best choice of location of a missing fragment is determined by the deepest hole in a pR1 map of a 3-dimensional location space. This best choice is combined with the trial orientation to form a candidate orientation-location. Trying all candidate orientations leads to a list of candidates of orientation-locations which can be ranked from smallest R1 to highest R1. The missing fragment is determined by the candidate orientation-location of the lowest R1. With this missing fragment being determined, a newer pR1 is defined by including the atoms of the newly determined fragment as known atoms. This newer pR1 is used to determine the next missing fragment in the same way.

When using single atom R1 (sR1) to locate single missing atoms (Zhang & Donahue, 2024), a rule excluding clustering ghost atoms and a rule excluding triangular bonding are enforced. Similarly, here, when using pR1 to locate a missing fragment, these rules are also enforced: if a trial orientation-location for a missing fragment will cause one of its atoms being a clustering ghost atom, or one of its atoms will involve triangular bonding with two known atoms, this trial is disqualified as a candidate for determining the orientation-location of the missing fragment.

## 4. Current implementation of using pR1 to determine missing fragments

In current implementation of using pR1 to determine missing fragments, the first step of detecting possible orientations of all missing fragments is same as described in above, but the second step of locating a missing fragment with a known trial orientation is slightly different. Instead of finding a set of pR1 holes and later determining which hole is the deepest one, the global minimum point of the pR1 is directly determined. This global minimum point is coarsely located with a grid of 0.4 Å within the cell: the grid point with the smallest R1 is the candidate global minimum point. Then the precision of locating this global minimum point is refined by locally halving the step size five times. To cut calculation time the coarse grid points are filtered by applying the rules excluding clustering ghost atoms and excluding triangular bonding. If after refinement the resulting global minimum point still passes these rules, it is accepted as the correct result, otherwise it is rejected, and the next best point of the coarse grid will be tried. This step will be repeated until there is an acceptable result.

## 5. Selected samples for test calculations

Four samples are selected and studied in this report. Table 1 lists the crystallography information of these samples.

**Table 1**   Crystallographic information of four selected samples

| Sample | Formula (excluding H) | Z | Non-hydrogen atoms in cell | a(Å) | b(Å) | c(Å) | α(°) | β(°) | γ(°) | space group |
|---|---|---|---|---|---|---|---|---|---|---|
| 1 | $S_2O_2C_{12}$ | 2 | 32 | 5.86 | 10.34 | 10.74 | 90 | 104.50 | 90 | P2(1) |
| 2 | $CuS_2P_2F_{12}N_8C_{18}$ | 2 | 86 | 10.93 | 14.67 | 8.14 | 90 | 90 | 90 | P2(1)2(1)2 |
| 3 | $C_{46}$ | 1 | 46 | 5.95 | 10.80 | 12.97 | 103.77 | 99.95 | 90.46 | P-1 |
| 4 | $C_{78}$ | 2 | 156 | 12.34 | 15.98 | 16.57 | 114.10 | 90.70 | 103.20 | P-1 |

Figure 2 shows the structures of these samples.

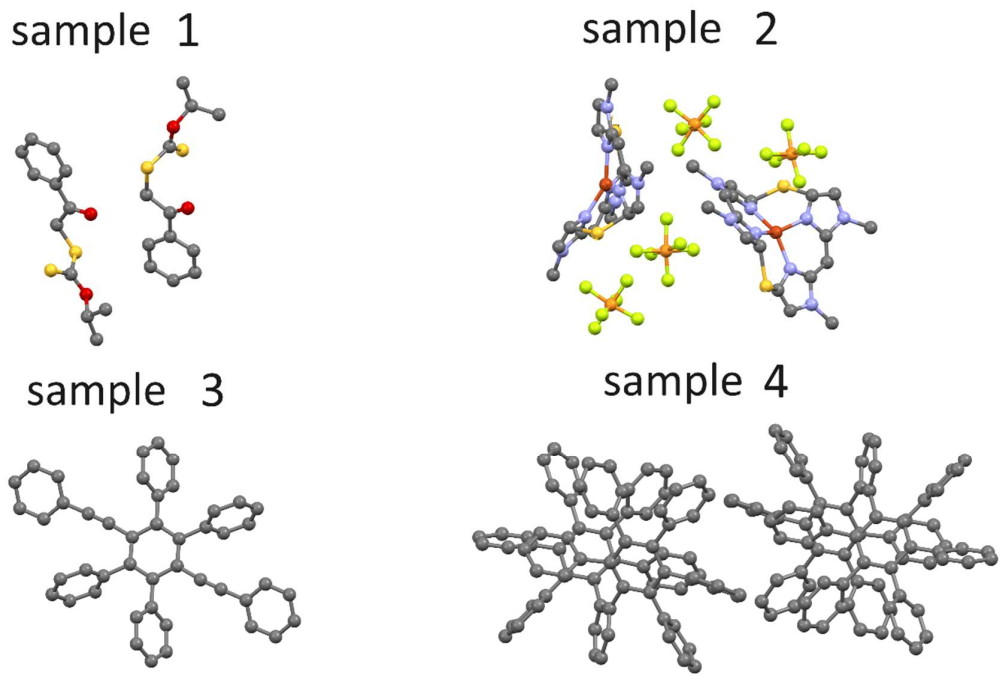

**Figure 2** The structures of the four samples studied in this report

**6. Test calculation with sample 1**

As seen in Figure 2, the structure of sample 1 has two $S_2O_2C_{12}$ molecules. The pR1 map of a free-standing $S_2O_2C_{12}$ molecule has 14340 holes, out of which 1000 deepest holes are selected for further study. These selected holes lead to 352 non-equivalent candidate orientations, which are ranked from smallest R1 to largest R1 and are labelled from 0 to 351. Because there are only two $S_2O_2C_{12}$ molecules, it is assumed that one molecule has orientation 0, and the other has orientation 1. Using orientation 0, molecule 0 is placed with its local origin at (0.3,0.3,0.3). A new pR1 is defined by treating the atoms of molecule 0 as the known atoms. Using orientation 1, molecule 1 is located by the global minimum point of this new pR1 map. Compared to the correct model, the final resulting model has all atoms being located within 0.5 Å.

**7. Test calculation with sample 2**

As seen in Figure 2, the structure of sample 2 has two Cu atoms, four $N_4C_9$ fragments, four $PF_6$ fragments, and four S atoms. The calculation starts by placing a single Cu atom at (0.3,0.3,0.3). The other Cu atom is then located by the deepest hole of an sR1 map. Next the four $N_4C_9$ fragments are considered. 1000 deepest holes of the pR1 map of a free-standing $N_4C_9$ fragment are determined. These holes lead to 272 non-equivalent candidate orientations, which are ranked from smallest R1 to largest R1 and are labelled from 0 to 271. Because there are only four $N_4C_9$ fragments, it is assumed that these fragments have orientations 0 to 3, respectively. With orientations of the $N_4C_9$ fragments being thus determined, each $N_4C_9$ fragment is located by the global minimum point of a pR1 map. Next, the $PF_6$ fragments are considered. There are 780 holes in the pR1 map of a free-standing $PF_6$ fragment. These holes lead to 18 non-equivalent candidate orientations, which are ranked from smallest R1 to highest R1 and are labelled from 0 to 17. As seen in Figure 2, the four $PF_6$ fragments share two orientations (each orientation is shared by a pair of $PF_6$ fragments). It is assumed (with a lucky guess) that these shared orientations are orientations 0 and 1. With their orientations being thus determined, the four $PF_6$ fragments are each located by the global minimum point of a pR1 map. Finally, the four S atoms are located one-by-one with sR1 calculations. The final model, if compared to the correct model, has all atoms located within 0.5 Å.

## 8. Predicting possible locations of all missing fragments

It is lucky that samples 1 and 2 only contain 2 to 4 fragments of each type of fragments. When there are only a few missing fragments, as practiced in above, it is possible to guess the orientations of the missing fragments. It is seen in Figure 2 that samples 3 and 4 are not so lucky: sample 3 has 7 benzene rings, and sample 4 has 26 benzene rings. When there are multiple missing fragments, it is hard to guess which candidate orientations should be used. In such situations, when locating a missing fragment, all candidate orientations should be tried. Each trial orientation yields a corresponding best choice of the location of the missing fragment. Trying all candidate orientations yields a list of candidate orientation-locations for the fragment. These candidate orientation-locations are ranked by R1. The one with the smallest R1 is used to determine the orientation and location of a missing fragment. Obviously, this type of strict calculation is very time consuming. Such strict calculation has been tried on sample 3 to locate its seven missing benzene rings, and the result is a complete success.

The holes of an sR1 map can predict possible locations of all missing atoms, a trick that has greatly shortened the calculation times in sR1 applications (Zhang & Donahue, 2024). Here in pR1 applications is there a way to predict the locations of all missing fragments? On the first glance, this is impossible because the fragments are of different orientations. But it becomes possible if a spherical symmetric model of a fragment is used. For any fragment, if all orientations are considered, taking average, the result is a completely disoriented spherical symmetric model. For this model, equation (1) is replaced by following two equations:

$F_c^2(hkl) \approx [f_1(hkl) \cos2\pi(hx_1+ky_1+lz_1) + \cdots + f_{i-1}(hkl) \cos2\pi(hx_{i-1}+ky_{i-1}+lz_{i-1})$

$\qquad + f_p(s) \cos2\pi(hx+ky+lz)]^2$

$\qquad + [f_1(hkl) \sin2\pi(hx_1+ky_1+lz_1) + \cdots + f_{i-1}(hkl) \sin2\pi(hx_{i-1}+ky_{i-1}+lz_{i-1})$

$\qquad + f_p(s) \sin2\pi(hx+ky+lz)]^2$

$\qquad + f_{j+1}^2(hkl) + \cdots + f_N^2(hkl)$ (2)

$$f_p(s) = [f_i(s)G(4\pi sr_i) + \cdots + f_j(s)G(4\pi sr_j)] \times n \qquad (3)$$

Here $n\equiv 1$, $s\equiv\sin\theta/\lambda$, $G(\chi)\equiv\sin\chi/\chi$, $f_j(s)\equiv f_j(hkl)$, $r_j$ is the distance from atom j to the local origin of the fragment, and $(x,y,z)$ is the location of the local origin of the fragment in the unit cell (in terms of fractional cell coordinates).

Using this model, a fragment is basically treated like a single atom. The possible locations of the local origin of all missing fragments are discovered as holes of a pR1 map of this spherical symmetric model.

By strict definition of a completely disoriented fragment, the parameter n in equation (3) equals 1. A model of disoriented fragment with n=1 can predict the locations of $PF_6$ fragments or benzene ring fragments. However, such a model has no prediction power when applied to $S_2O_2C_{12}$ molecule or $N_4C_9$ fragment. The main reason for this failure is because the space occupied by a disoriented model of one of these fragments overlaps the nearby other atoms. Interestingly, in these cases, if n is increased to 1000, the prediction power resumes. However, in the following applications of predicting the locations of benzene rings, n=1 will be used.

**9. Test calculation with sample 3**

The structure of sample 3 has 7 benzene rings. The pR1 map of a free-standing benzene ring of sample 3 has 4176 holes. These holes lead to 35 non-equivalent candidate orientations, which are ranked from smallest R1 to largest R1 and are labelled from 0 to 34. Using orientation 0, benzene ring 0 is placed with its local origin at (0.3,0.3,0.3). A new pR1 is defined by treating the atoms of benzene ring 0 as the known atoms. The possible locations of the local origins of all missing benzene rings are predicted by the holes of a pR1 map of the completely disoriented benzene ring model. About a few hundred possible locations are predicted. To determine benzene ring 1, all 35 candidate orientations are tried. For each trial orientation, R1 are calculated at the predicted locations (after enforcing rules excluding both clustering ghost atoms and triangular bonding). The location with the lowest R1 is the best choice, which is combined with the trial orientation to form one candidate orientation-location. After trying all candidate orientations, a list of candidate orientation-locations are formed. From this list the one with the lowest R1 is used to determine the orientation and location of benzene ring 1. Then a newer pR1 is defined by accepting the atoms of benzene ring 1 as the known atoms, and the same calculations are repeated to determine benzene ring 2. Similarly, benzene rings 3, 4, 5, and 6 are determined. Upon a visual inspection of the resulting model, all benzene rings are relatively placed in ways that are chemically meaningful, except benzene ring 0. Note that there are four missing C atoms. Benzene ring 0 is located such that no matter how to add the missing C atoms it cannot be connected to the main structure in a chemically meaningful way. Benzene ring 0 is deleted and re-discovered by resuming the calculations. Finally, the remaining four missing C atoms are determined by sR1 calculations. The final model, when compared to the correct model, has all atoms located within 0.5 Å.

## 10. Test calculation with sample 4

The structure of sample 4 has 26 benzene rings. The pR1 map of a free-standing benzene ring of sample 4 has 432 holes. These holes lead to 29 non-equivalent candidate orientations, which are ranked from smallest R1 to largest R1 and are labelled from 0 to 28. The determination of the missing benzene rings goes in the same way as done in sample 3. In the end, benzene rings 0 to 25 are all determined. This result is checked visually, and the suspected bad benzene rings are identified. If a ring is isolated, without bonding to anything, it is suspected to be bad. Some benzene rings are bonded to other benzene rings in chemically incorrect ways. They are also suspected to be bad ones. When two benzene rings are connected by a single bond correctly, this bond is aligned with the centers of both rings. Then how to identify a bad benzene ring? If benzene rings A and B are bonded correctly, but A and C are bonded incorrectly, then C is suspected to be a bad one. In another situation, if several benzene rings are bonded to benzene ring O, and they are all bonded to O in chemically wrong ways. Upon a closer inspection, it is seen that if benzene ring O is oriented slightly differently, all the bonds will be chemically correct. In this case, benzene ring O is suspected to be a bad one. Applying these judgements, 12 benzene rings are suspected to be incorrectly determined. After deleting these suspected bad benzene rings, the calculation is resumed. Inspection of the new result reveals that benzene rings 0 to 24 are bonded to each other in chemically correct ways, however, benzene ring 25 is isolated without bonding to anything. When determining benzene ring 25, orientation 16 gives the lowest R1 of 0.5827, however, its resulting benzene ring is isolated. Upon investigation, the second-best orientation is orientation 2, which has the second lowest R1 of 0.5836. This orientation is initially rejected because it has a slightly higher R1. Because this orientation yields chemically correct benzene ring, the result of this orientation should be adopted. With this correction, the final model, if compared to the correct model, has all atoms located within 0.5 Å.

## 11. General strategy of using pR1 in MR calculations

The possible orientations of all missing fragments of one type are detected by the holes in a pR1 map of a free-standing fragment in a 3-dimensional orientation space. Non-equivalent orientations are filtered out and ranked from the smallest R1 to the highest R1. To strictly determine the orientation and location of a missing fragment, all candidate orientations should be tried. The best location corresponding to each trial orientation is determined by the deepest hole of a pR1 map in a 3-dimensional location space. By trying all candidate orientations, a list of orientation-best-locations are formed, from which the one with the smallest R1 determines the orientation and location of a missing fragment. To cut calculation time, the possible locations of all missing fragments can be predicted as the holes in a pR1 map of a completely disoriented fragment. The calculation time can be shortened because for each trial orientation, only these predicted locations need to be tested for locating the global minimum point of a pR1 map. Finally, it is necessary to examine the resulting model for possible errors. The fragments that are bonded in chemically incorrect ways should be deleted and re-discovered. The pR1 method can improve model quality, just like sR1 method; the reason is the same: pR1 has improved precision in locating missing fragments when the model becomes more complete.

## 12. Why is pR1 a successful MR target

The traditional R1 target was only occasionally used as MR target (Eventoff *et al.*, 1975; Bott & Sarma, 1976), and it is generally regarded as not a good MR target (McCoy, 2017). However, this report has demonstrated that pR1 is a successful MR target. How can all these observations be reconciled with each other? The key to this question is the difference between an R1 target and the pR1 target. To explain this difference, the atoms of the target structure should be divided into three groups: (1) the known atoms 1 to i-1, (2) the missing atoms i to j of a missing fragment, and (3) the other missing atoms j+1 to N. The atoms of group (1) and (2) make up a current working model. Therefore, an R1 target uses an equation like equation (1) but without the tail $f_{j+1}^2(hkl) + \cdots + f_N^2(hkl)$. On the other hand, pR1 uses equation (1) with this tail retained. The locations of the other missing atoms j+1 to N do not appear in an R1 target; they do not appear in pR1, either. However, pR1 has the part $f_{j+1}^2(hkl) + \cdots + f_N^2(hkl)$ retained. This part accounts the baseline effect of these missing atoms on the R1. By retaining this part, pR1 is calculated as accurate as possible. Because a single fragment has weak effect on R1, if this baseline effect of the other missing atoms is not accounted for, pR1 will turn into the R1 target which is a worse MR target. The maximum likelihood target (Bricogne, 1992; Read, 2001; McCoy, 2004) is an improvement over the R1 target in terms of a rigorous statistical basis, but it inherits the drawback of the R1 target by also ignoring the effects of the other missing atoms. In comparison, pR1 target loses the battle of seeking a rigorous statistical basis, but it wins the battle of accounting the baseline effect of the other missing atoms. The LLGI target (McCoy *et al.*, 2007; Read & McCoy, 2016) mitigates this problem by only taking the "gain" as signal: though both the model and the non-informative model ignore the baseline effect of the other missing atoms j+1 to N, this ignored effect is cancelled out (to some extent) when the difference (the gain) between the two models are calculated, so, the final signal is purely due to the useful information of the model. Currently, pR1 is only practiced with a few examples, so it is premature to speculate how pR1 can compete with the well-established LLGI target.

## 13. Future directions

In current implementation, pR1 calculations do not require prior knowledge of the symmetry of the target structure. When such prior knowledge does exist, it can be exploited to shorten the calculation time. As how this can be done is something that needs to be studied in future.

This report only has tried pR1 calculations on four small structures. To extend pR1 calculations to macromolecules is a subject that needs to be investigated in future.

**Acknowledgements**     Professor Robert A. Pascal has kindly provided the data sets of samples 18 and 20.

# Supporting information

### S1. Information on data collection of the samples

All crystals were coated with paratone oil and mounted on the end of a nylon loop attached to the end of the goniometer. Data were collected at 150 K under a dry $N_2$ stream supplied under the control of an Oxford Cryostream 800 attachment. The data collection instrument was a Bruker D8 Quest Photon 3 diffractometer equipped with a Mo fine-focus sealed tube providing radiation at $\lambda = 0.71073$ nm or a Bruker D8 Venture diffractometer operating with a Photon 100 CMOS detector and a Cu Incoatec I microfocus source generating X-rays at $\lambda = 1.54178$ nm.

### S2. Choice of reference vectors r and s for each fragment

The labels of the atoms are read from Figure 1 in the body of the report.

For $S_2O_2C_{12}$ molecule: reference vector **r** is the vector from C10 to C11, and reference vector **s** is the vector from C10 to C9.

For $N_4C_9$ fragment: Reference vector **r** is the vector from N2 to N3, and reference vector **s** is the vector from N2 to C12.

For $PF_6$ fragment: Reference vector **r** is the vector from P1 to F2, and reference vector **s** is the vector from P1 to F4. The P-F bond length is 1.59 Å.

For benzene ring fragment: let O be the center of the benzene ring. Reference vector **r** is the vector from O to C1, and reference vector **s** is the vector from O to C2. The C-C bond length of a benzene ring is 1.39 Å. For benzene fragment, the relations between the unit vectors **x**, **y**, and **z** and the reference vectors **r** and **s** are: $\mathbf{y} = \mathbf{r}/|\mathbf{r}|$, $\mathbf{z} = (\mathbf{s}-\mathbf{s}\cdot\mathbf{yy})/|\mathbf{s}-\mathbf{s}\cdot\mathbf{yy}|$, $\mathbf{x} = \mathbf{y}\times\mathbf{z}$.

### S3. Mathematical formula describes a general rotation

Before rotation, a point has Cartesian coordinates $(x,y,z)$. After a rotation of angles $(\psi,\varphi,\zeta)$, the point moves to a new location in the same Cartesian system with Cartesian coordinates $(x',y',z')$, which are calculated by:

$$\begin{pmatrix}x'\\y'\\z'\end{pmatrix} = \begin{pmatrix}1 & 0 & 0\\ 0 & \cos\psi & -\sin\psi\\ 0 & \sin\psi & \cos\psi\end{pmatrix}\begin{pmatrix}\cos\varphi & -\sin\varphi & 0\\ \sin\varphi & \cos\varphi & 0\\ 0 & 0 & 1\end{pmatrix}\begin{pmatrix}1 & 0 & 0\\ 0 & \cos\zeta & -\sin\zeta\\ 0 & \sin\zeta & \cos\zeta\end{pmatrix}\begin{pmatrix}x\\y\\z\end{pmatrix}$$

In general, to cover all possible rotations, the range of $\psi$ should be 0 to 360 degrees, the range of $\varphi$ should be 0 to 180 degrees, and the range of $\zeta$ should be 0 to 360 degrees. When a fragment has n-fold rotation symmetry and the rotation axis is along z-axis of its local Cartesian system, then the range of $\zeta$ can reduce to 0 to 360/n degrees.

**S4. Mathematical formula relating the cell fractional coordinates and the cell Cartesian coordinates**

Using cell parameters $a$, $b$, $c$, $\alpha$, $\beta$, and $\gamma$, the relation between cell fractional coordinates $(x_f, y_f, z_f)$ and cell Cartesian coordinates $(x_c, y_c, z_c)$ is expressed as:

$$\begin{pmatrix} x_c \\ y_c \\ z_c \end{pmatrix} = \begin{pmatrix} 1 & b\cos\gamma & c\cos\beta \\ 0 & b\sin\gamma & \dfrac{c(\cos\alpha - \cos\beta\cos\gamma)}{\sin\gamma} \\ 0 & 0 & \dfrac{V}{ab\sin\gamma} \end{pmatrix} \begin{pmatrix} x_f \\ y_f \\ z_f \end{pmatrix}$$

in which the cell volume V is calculated as:

$$V = abc\sqrt{1 - \cos^2\alpha - \cos^2\beta - \cos^2\gamma + 2\cos\alpha \times \cos\beta \times \cos\gamma}$$